\documentclass[prd,letterpaper,nofootinbib,twocolumn,superscriptaddress,aps]{revtex4}
\usepackage[dvipsnames]{xcolor}
\usepackage{fancyhdr}
\usepackage{float}
\usepackage{listings}
\usepackage{enumerate}
\usepackage[T1]{fontenc}
\usepackage{ae}
\usepackage{rotating}
\usepackage{subfigure}
\usepackage{amsmath}
\usepackage{amssymb}
\usepackage{graphicx}
\usepackage{dcolumn}
\usepackage{bm}
\usepackage{hyperref}
\usepackage{slashed}
\usepackage[sort&compress]{natbib}
\usepackage[normalem]{ulem}
\usepackage{multirow}
\usepackage{mathrsfs}
\usepackage{lipsum}

\newcommand{\sylm}{{}^{-2}Y_{lm}(\iota,\phi)}

\begin{document}
\title{Enhancing the Sensitivity of Searches for Gravitational Waves from Core-Collapse Supernovae with a Bayesian classification of candidate events}
    
\newcommand*{\ER}{Embry Riddle University, 3700 Willow Creek Road, Prescott Arizona, 86301, USA}
\affiliation{\ER}
\newcommand*{\UTGRV}{The University of Texas Rio Grande Valley, One West University Boulevard, Brownsville, 78520, USA}
\affiliation{\UTGRV}
\author{K. Gill} \affiliation{\ER}
\author{W. Wang} \affiliation{\UTGRV}
\author{O. Valdez} \affiliation{\UTGRV}
\author{M. Szczepa\'nczyk}  \affiliation{\ER}
\author{M. Zanolin} \affiliation{\ER}
\author{S. Mukherjee} \affiliation{\UTGRV}
\begin{abstract} 
We demonstrate how a morphological veto involving Bayesian statistics can improve the receiver-operating characteristic (ROC) curves of the current search for core-collapse supernovae (CCSNe) as implemented by the coherent Waveburst (cWB) algorithm. Examples involving two implementations of BayesWave (BW), one that makes no assumption of the polarization state of the gravitational wave (GW) and one that uses the same elliptical polarization settings adopted in previous usages for Binary systems are provided on the set of waveforms currently adopted for the first and second Advanced LIGO (aLIGO) science runs for the targeted CCSNe search. A comparison of the performance for all-sky searches versus the targeted searches with optical triggers is provided. The average cWB+BW ROC range improvements with respect to a fixed false-alarm rate (FAR) for slowly-rotating waveforms ranged from [$+1.30\%$, $+15.76\%$] while the improvement for rapidly-rotating waveforms were on the order of [$+1.19\%$, $+22.05\%$]. The application of BW to CCSN GW triggers also shows a significant reduction of the FAR while maintaining detection efficiency and remaining sensitive to a wide range of morphological CCSNe signals. It appears that the code developed for arbitrarily polarized signals outperforms the previous code for the GW morphologies tested.
\end{abstract}
\maketitle

\section{Introduction}

\indent A central challenge in GW astronomy with laser interferometers is distinguishing weak signals from transient detector artifacts (glitches)\,\cite{2016CQGra..33m4001A}. Detections of GW signals have the potential to test and constrain emission models of astrophysical sources\,\cite{2016arXiv160501785A, 2016PhRvX...6d1015A, 2017ApJ...841...89A}. This requires reconstructing the signal waveform from the GW detector output, estimating parameters of the waveform and applying selection cuts to eliminate as many noise-induced candidates\,\cite{2016arXiv160501785A, 2016PhRvX...6d1015A, 2017ApJ...841...89A} as possible. Data analysis for binary systems strongly benefits from having deterministic templates, which allows the use of matched filtering for detection purposes. CCSNe simulations, however, have not reached the same confidence to claim robust estimates of the signals as the evolution itself is turbulent and the signals from CCSNe are expected to be weaker\,\cite{doi:10.1146/annurev.nucl.55.090704.151608, 2016PASA...33...48M, 2014ApJ...786...83T}. Therefore, methods that improve the sensitivity of searches for GW signals from CCSNe are desirable in the advanced detector era and beyond.

\indent The goal of this paper is to customize the procedure of calculating the Bayesian likelihood ratios for CCSNe in a way such that noise events and events produced by CCSNe GWs may be separated with benefits to both the detection confidence and detectability range.

\indent An important advantage in searching for CCSNe waveforms in the presence of multi-messenger observations is the capability to narrow down the interval of time when the GW transits the earth (on-source window) because it allows to detect GWs further from the Earth without increasing the false alarm probability (FAP). The set of thresholds that both cWB and BW adopt depend on the desired FAR, where the FAR is the FAP divided by the on-source window, and the pool of waveforms over which the performance is desired to be robust. 

\indent In order to quantify the improvements of cWB+BW versus cWB alone, in this paper we produce ROCs in different regimes of FAP and FAR of interest for different on-source windows (Fig 8 allows to identify the FAR of interest depending on the duration of the on-source window and the desired number of sigma in the detection confidence). 

\indent In this analysis, we focus on distances where either cWB alone or cWB+BW reach a $50\%$ detection efficiency for current CCSNe waveforms. It should be stressed that BW is a follow up module, and can only be used to veto noise events (or reconstruct their properties). BW cannot detect potential events that are not identified by cWB. Accordingly, the cWB+BW operating point, is always located at lower or equal FAR and detection efficiencies. This does not mean, however, that the cWB+BW ROC curve does not stand above the one for cWB alone, and in this paper we actually prove that it does. We use two versions of BW: one for linearly polarized, and one for arbitrary polarized waveforms. We also employ a portion of S5 data, which is mostly, but not only, dominated by Gaussian noise events. The choice of S5 data follows that it is the only data set where CCSNe results are published with a sizable coverage of the on-source window. Furthermore the choice of a set of data with little non-Gaussian events can provide a conservative estimate of the benefits of BW, since non-Gaussian glitches were morphologically different with respect to events induced by GWs. Also, it is expected that non-Gaussian glitches evolve from one science run to the next and the exact results might have less general value. The analysis in this paper also shows that the methodology that BW adopts to compute the probability of an event being a signal versus just noise is not as accurate as in the previous applications for compact binary coalescence, CBC, signals and might require further investigations for CCSNe. Overall, however, the usage of BW appears to be promising either for FAR reduction (especially if the cWB ROCs are flat) or efficiency improvements.

\indent The main conclusion of this analysis is that the newly introduced arbitrarily polarized code consistently improves either the detection efficiency or detection confidence. An improvement in efficiency at fixed FAR could be corrected to an improvement in range by seeing which range rescaling in the efficiency curve produces the observed efficiency improvement in the ROC. For example, a $30\%$ efficiency improvement at a $50\%$ cWB detection efficiency corresponds to a $30\%$ improvement in the range.

\indent The paper is organized as follows. In section II, we introduce the coherent WaveBurst (cWB) pipeline\,\cite{2008CQGra..25k4029K}, the BayesWave (BW) algorithm\,\cite{2015CQGra..32m5012C, 2016arXiv161202003B} and the improvement metrics. In section III, we review waveforms from multidimensional, magnetohydrodynamical simulations used, including the different GW polarizations expected from CCSN signals and discuss the differences in searching for CCSNe signals with a fixed direction or sky mask like in the initial LIGO search\,\cite{2016arXiv160501785A}. In section IV, we discuss the results and implications of this analysis.

\section{Methodology}

\indent In this paper, we quantify the performance improvement brought by BW to cWB with the following metrics: \textit{(a)} improvement of the detection efficiency at a fixed FAR or FAP and \textit{(b)} the reduction of FAR/FAP at fixed detection efficiencies. We also discuss how the second metric is the main benefit in the presence of flat ROC curves. 

\indent This study uses, as an example, an on-source window with S5 data for a two-detector network, LIGO Hanford H1 and Livingston L1. The data that is analyzed in this paper is the data used for the initial LIGO/Virgo SN search with SN2007gr, with 5.11 days of on-source window and 3.71 days of coincident data where both detectors were locked. The noise spectral density in these 3.71 days of coincident data is about 30 times worse than the projected Advanced LIGO sensitivity and a factor of about 300 to the proposed Cosmic Explorer configuration\,\cite{2017CQGra..34d4001A}. cWB ranks the triggers with network signal-to-noise\,\cite{2012JPhCS.363a2032N} denoted as $\rho$. We produce a list of noise triggers in time-shifted data to establish the FAR with respect to $\rho$ (Fig 1). A large FAR corresponds to a lower threshold for $\rho$. We use BW as a follow-up to remove as many as possible noise-induced cWB events. 
\begin{figure}[h!] 
   \centering
   \includegraphics[width=3.4in]{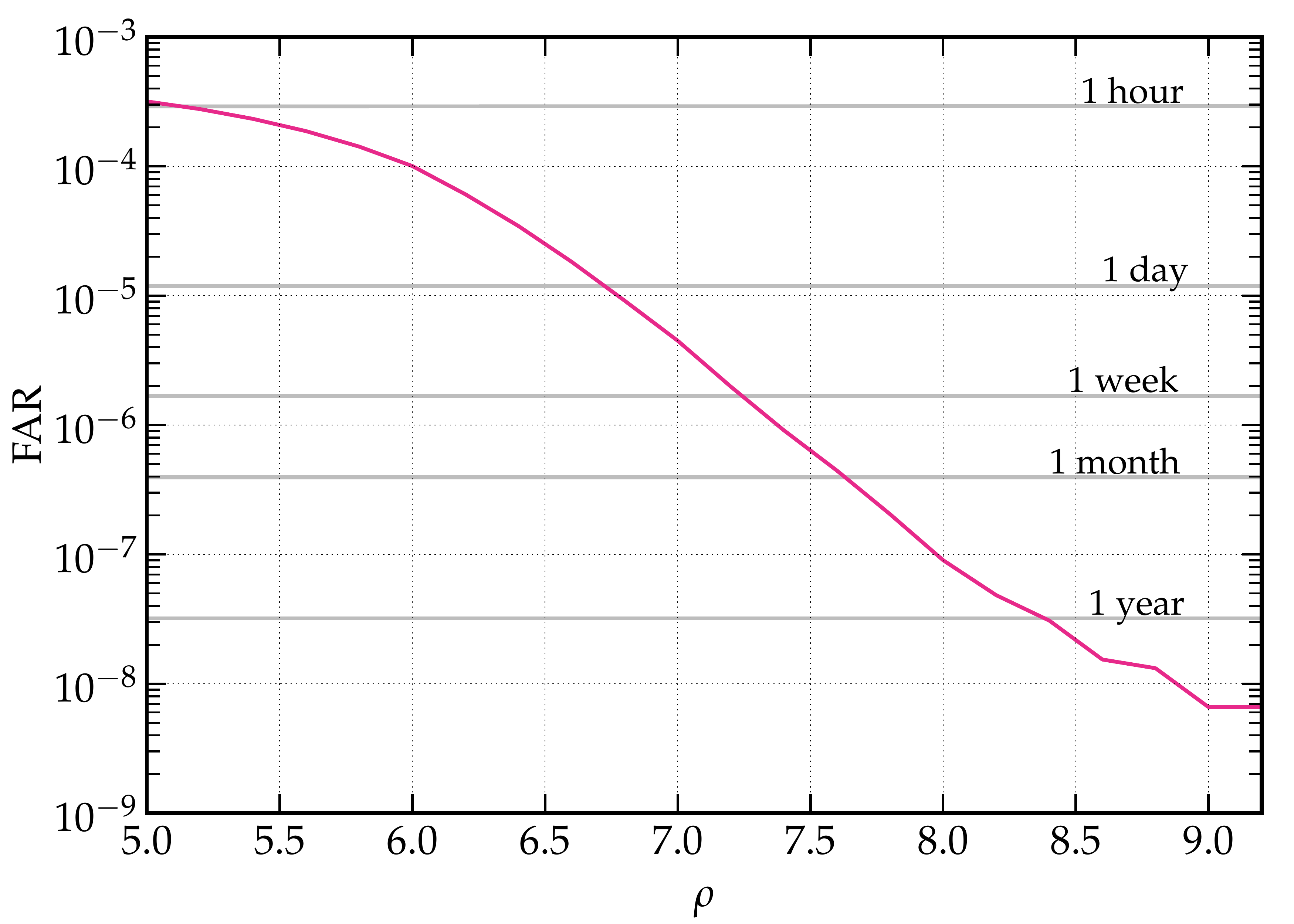}
   \parbox{3.4in}{\caption 	{FAR with respect to $\rho$ for different on-source windows for the 3.71 days of coincident data of S5 used in this paper. A large FAR corresponds to a lower threshold for $\rho$. We use BW as a follow-up to remove as many as possible noise-induced cWB events. The improvement in ROCs are due to having a lower threshold for $\rho$ for the same final FAR with the combination of cWB+BW.}}
   \label{Figure 1}
\end{figure}

\subsection{Coherent WaveBurst (cWB)} 
\indent Coherent WaveBurst\,\cite{2008CQGra..25k4029K} (cWB) is an excess power pipeline currently designed to search for GW bursts of the order of a second long. The pipeline decomposes the detector data using WDM wavelets\,\cite{2012JPhCS.363a2032N} and identifies time-frequency regions with coincident energy between detectors and requires some degree of coherence. The algorithm chooses $0.2\%$ of the time-frequency regions for further analysis. The triggers are chosen based on the excess power that is coincident between the detectors without any other prior assumption.

\indent In order to attach significance to the GW candidate, the background analysis needs to be performed through time shifting the detector stream from one detector with respect to the other detector. It allows to understand how often the detectors noise can create a trigger that looks like a GW. The background analysis establishes the cWB detection statistic, $\rho$,  derived from a constrained maximum likelihood framework, and asymptotically reproduces the matched filter SNR for large amplitudes (for example, Fig 2). In this work, the data was shifted 300 times that gives total of $\sim$3 years of background data and searched for triggers that fall in the time-frequency range between (64, 2048) Hz and are resampled to 4096 Hz. This allows for the pipeline to capture majority of the GW energy in the signal as well as limiting the computational cost.

\begin{figure}[b!] 
   \centering
   \includegraphics[width=3.4in]{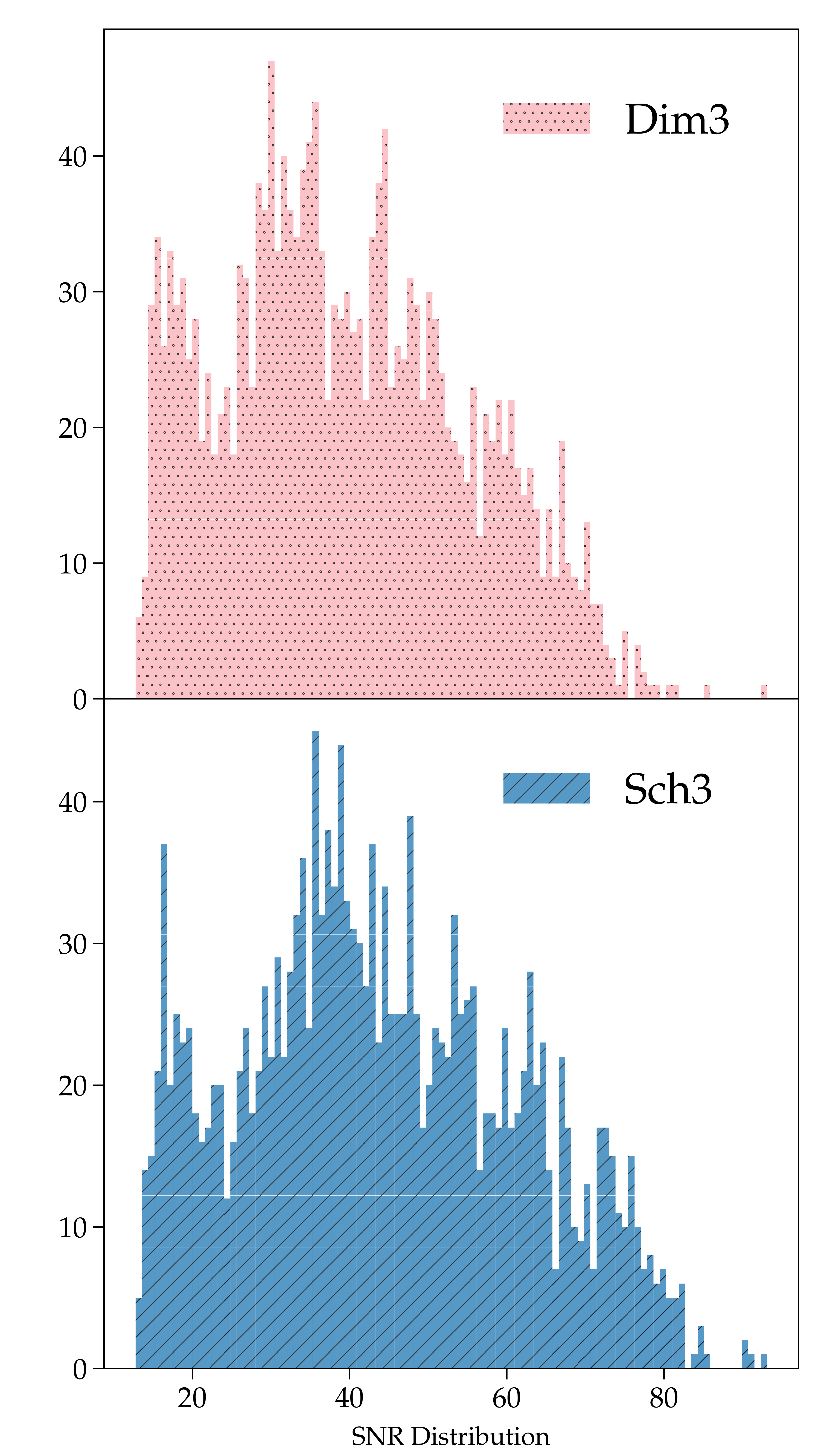}
   \parbox{3.4in}{\caption 	{Reconstructed cWB injected SNR amplitudes at $50\%$ efficiency for \textbf{dim3} and \textbf{sch3} at a nominal distance of 10 kpc.}}
   \label{Figure 1}
\end{figure}

\subsection{BayesWave (BW)} 
\indent The BW algorithm reconstructs signals and glitches using a linear combination of sine-Gaussian wavelets, where the number of wavelets needed in either the glitch or signal model is optimized using a reversible jump Markov chain Monte Carlo. The glitch model uses the data separately in each interferometer with an independent linear combination of wavelets. The signal model reconstructs the candidate event at some fiducial time corresponding to the center of the Earth, taking into account the response of each detector in the network to that signal. BW also uses a parameterized phenomenological model, BayesLine, to assume Gaussian noise in modeling the instrument noise spectrum\,\cite{2015PhRvD..91h4034L, 2016PhRvD..93l2004A}.

[floatfix]
\indent For each detection candidate, BW computes \textit{Bayes factors}\,\cite{sivia2006data}, $\mathrm{B_{LN(S/N)}}$ as its discrimination statistic. This allows BW to assign a higher statistic to signals possessing non-trivial time-frequency structures as well as providing a measure of the signal complexity. 

\indent Bayes factors are reported in a natural logarithmic scale, lnBSG, for which the number of wavelets used in the reconstruction appears to scale with the log of the SNR. The $\mathrm{B_{LN(S/N)}}$ is the ratio of the estimated probability of having a GW transient given the collected data (hypothesis $\mathrm{S_0}$) with the estimated probability that only Gaussian noise is present in the data (hypothesis $\mathrm{N_0}$).

\begin{equation}
\mathrm{B_{LN(S/N)}} = \frac{p(S_0|x, I)}{p(N_0|x, I)} = \frac{p(S_0| I)}{p(N_0| I)} \frac{p(x|S_0, I)}{p(x|N_0, I)}
\end{equation}

\indent where $x$ is the data, $I$ is the set of shared assumptions, such as detector locations, orientations, noise power spectra. In the presence of a GW signal, we expect larger probability ratios while smaller values indicate noise. The overlap of values of $\mathrm{B_{SN}}$ for the two hypothesis depends on the accuracy of the estimation of the probabilities. The probabilities in Eqn (1) are computed assuming a gaussian distributed noise time-series with isolated non-Gaussian glitches. The reconstructed signal comes from performing a wavelet decomposition\,\cite{2015CQGra..32m5012C,2013PhDT.......248B,2016arXiv161202003B}. If the probability estimates are unbiased, there is no overlap and the separation at $\mathrm{B_{SN}}$ = 1. BW also allows to calculate a signal-to-non-Gaussian glitch ratio ($\mathrm{B_{SG}}$). The logarithms of these ratios allow the production of scatter plots for the populations of noise and GW events (for example, Figure 9-12) where we indicate a separation line (eqn (12)), which allows a consistent interpretation of the ROCs, used in the tuning for this paper (table IV).

\indent It is important to stress that the usefulness of this separation for CCSNe GW ROC improvements does not mean that all the assumptions hold robust. Figures 9-12 actually indicate that more work is necessary to understand how to reliably estimate these probabilities for CCSNe GWs.

\begin{table*}[t]
\caption{\label{tab:table2}SN waveform families along with cWB distance measurements at $50\%$ detection efficiency at $\mathrm{10^{-6}}$ FAR ($\rho$ threshold of 7.4 for S5 data).}
\begin{ruledtabular}
\begin{tabular}{c|c|c|c}
Emission Type & Waveform Identifier & Polarization & $50\%$ cWB Distance (kpc)\\ [0.5ex] 
\hline 
2D Rotating Core Collapse (s15A2O05ls) & dim1 & + & 0.36 \\
2D Rotating Core Collapse (s15A2O09ls) & dim2 & + & 0.42 \\
2D Rotating Core Collapse (s15A2O15ls) & dim3 & + & 1.10\\
3D Neutrino-Driven Convection and SASI & ott & +, $\times$ & 0.12 \\
3D Rotating Core Collapse & sch1 & +, $\times$ & 2.28\\
3D Rotating Core Collapse & sch2 & +, $\times$ & 1.87 \\
3D Rotating Core Collapse & sch3 & +, $\times$ & 1.37 \\
2D Neutrino-Driven Convection and SASI (B12-WH07) & yak1 & + & 2.28 \\ 
2D Neutrino-Driven Convection and SASI (B15-WH07) & yak2 & + & 3.11 \\ 
2D Neutrino-Driven Convection and SASI (B20-WH07) & yak3 & + & 3.06 \\ 
2D Neutrino-Driven Convection and SASI (B25-WH07) & yak4 & + & 5.25 \\ 
\end{tabular}
\end{ruledtabular}
\end{table*}

\section{CCSNe Emission models and their GW Signatures}

\indent In this section, we summarize the waveforms adopted in this study and the polarization states expected to emerge (table I).

\indent \textit{Non-rotating Core Collapse driven by Convection and Standing Accretion Shock Instability}: Ott et al. \,\cite{2013ApJ...768..115O} is a general-relativistic 3D simulation incorporating a three-specific neutrino leakage scheme producing both polarizations. We used the GW waveform, \textbf{ott}, from model s27fheat1.05 (a 27 $\mathrm{M\odot}$ progenitor). Yakunin et al. \,\cite{PhysRevD.92.084040} uses 2D hydrodynamics with relativistic corrections and a Lattimer-Swesty nuclear equation of state with a nuclear incompressibility, K = 220 MeV for a set of nonrotating Woosley-Heger, nonperturbed axisymmetric progenitor models. We use the GW waveform, \textbf{yak}, from model B15-WH07 (a 15 $\mathrm{M\odot}$ progenitor).

\indent \textit{Rotating Core Collapse Bounce dominated GW}: If the stellar core is rotating, the dynamics of the evolution are reflected in the GW signal waveform. The waveform encompasses a slow rise during core collapse, a large negative peak around core bounce, and damped oscillations experienced during the ring-down phase. Rotational effects also have the possibility of stopping core collapse due to centrifugal forces at subnuclear density. For very rapid rotation\,\cite{2008PhRvD..78f4056D,2012PhRvD..86b4026O,2014PhRvD..89d4011K}, the peak of the GW amplitude reaches a maximum and declines again while the centrifugal barrier considerably slows down the contraction and prevents the core from collapsing to higher super nuclear densities, with strongly rotating progenitors to possess initial axisymmetric dynamics and a linearly polarized emission over the same period.

\indent We selected three representative 2D rotating core collapse waveforms from Dimmelmeier et al.\,\cite{2002A&A...393..523D}, illustrating that the signal waveform remains qualitatively unaltered for a wide range of initial rotational strength using a 15$\mathrm{M_{\odot}}$ progenitor star and the Lattimer-Swesty nuclear equation of state with a nuclear incompressibility, K = 180 MeV\,\cite{1991NuPhA.535..331L}. \textbf{dim1} is fixed with moderate rotation while \textbf{dim3} is associated with extreme rapid rotation.

\indent Using 3D Newtonian, magnetohydrodynamical simulations that were based on a neutrino leakage scheme, Scheidegger et al.\,\cite{2010A&A...514A..51S} simulations were performed using a 15$\mathrm{M_{\odot}}$ progenitor star and the Lattimer-Swesty equation of state with K = 180 MeV\,\cite{1991NuPhA.535..331L}. \textbf{sch1}, the $\mathrm{R1E1CA_L}$ waveform model, employs slow pre-collapse rotation with a toroidal/poloidal magnetic field with the strength of $10^{9}$ G/$10^{6}$ G. \textbf{sch2}, the $\mathrm{R3E1AC_L}$ waveform model, moderate pre-collapse rotation with a toroidal/poloidal magnetic field with the strength of $10^{6}$ G/$10^{9}$ G. \textbf{sch3}, the $\mathrm{R4E1FC_L}$ waveform model, is a rapid pre-collapse rotation with a toroidal/poloidal magnetic field strength of $10^{12}$ G/$10^{9}$ G.

\begin{figure}[b] 
   \centering
   \includegraphics[width=3.4in]{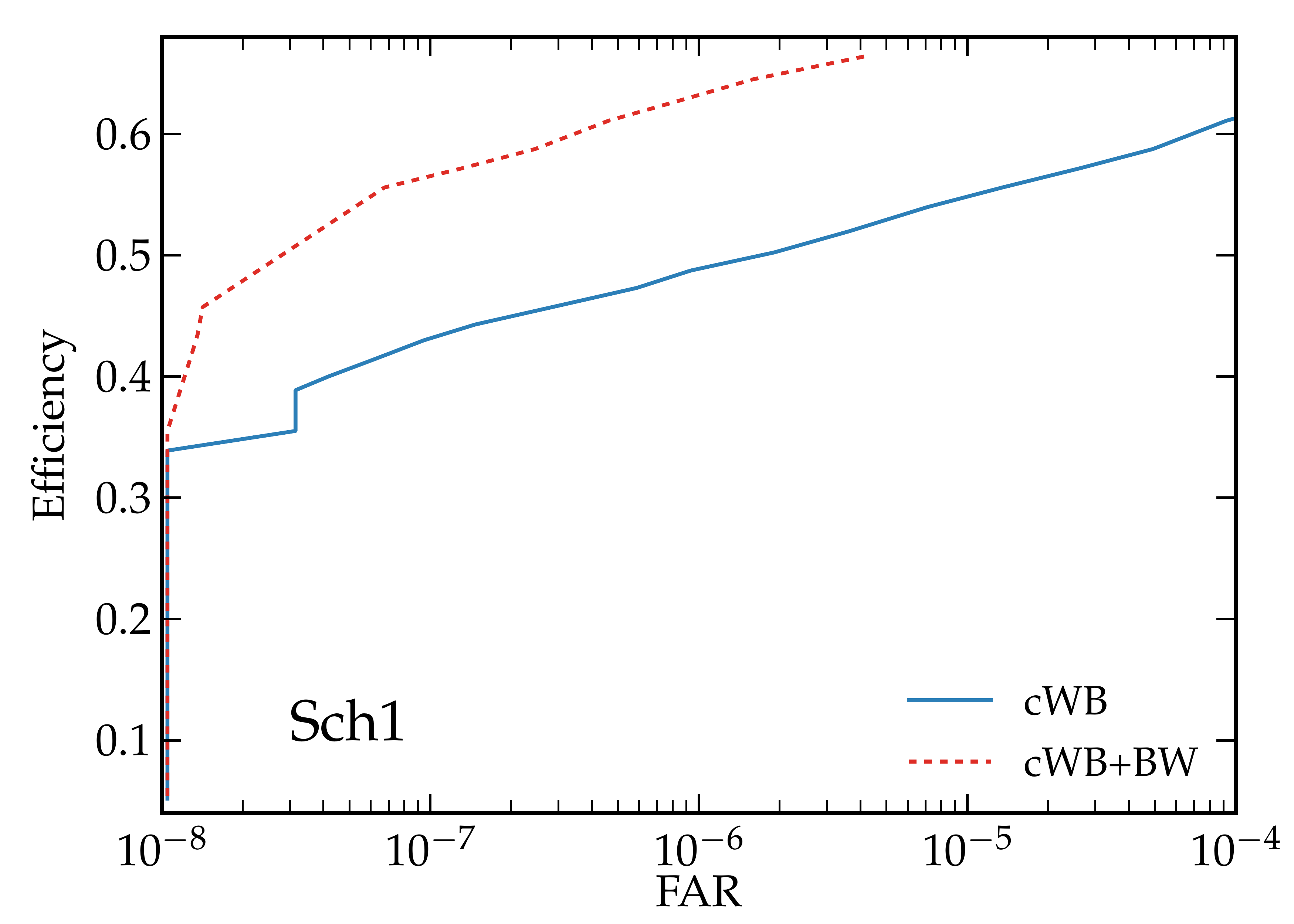}
   \parbox{3.4in}{\caption 	{ROC Improvement using a combination of cWB+BW for \textbf{sch1} injected at a nominal distance of 10 kpc with a $50\%$ cWB detection efficiency distance of 2.28 kpc. The plot illustrates both an efficiency improvement at fixed FAR and that the $50\%$ efficiency is achieved at a FAR two orders of magnitude smaller.}}
   \label{Figure 1}
\end{figure}

\begin{figure}[h!] 
   \centering
   \includegraphics[width=3.4in]{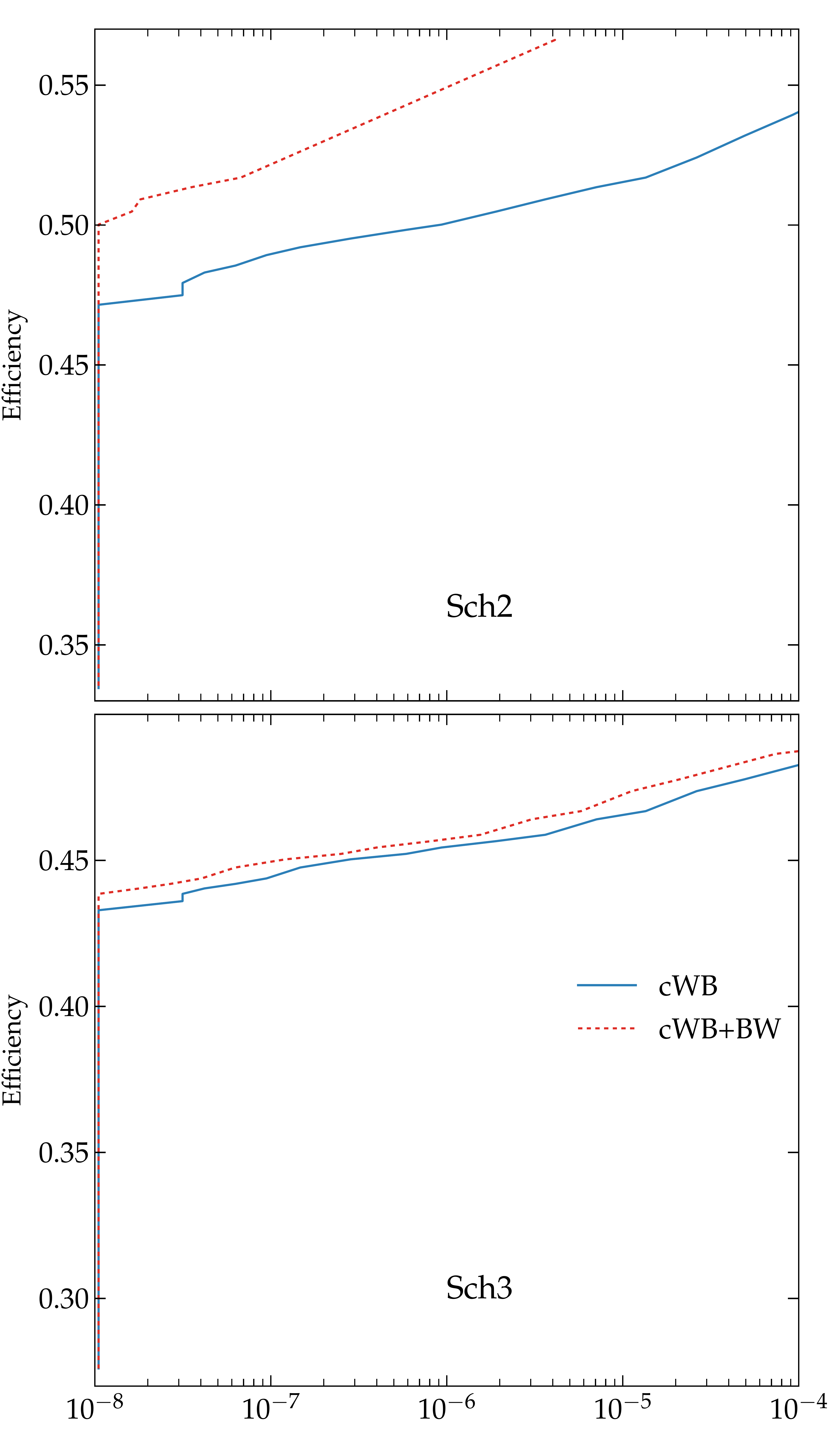}
   \parbox{3.4in}{\caption 	{ROC Improvement using a combination of cWB+BW for \textbf{sch2/3} injected at a nominal distance of 10 kpc with a $50\%$ cWB detection efficiency distance of 1.87 kpc (\textbf{sch2}) and 1.37 kpc (\textbf{sch3}).\color{black}}}
   \label{Figure 1}
\end{figure}

\begin{figure}[h] 
   \centering
   \includegraphics[width=3.4in]{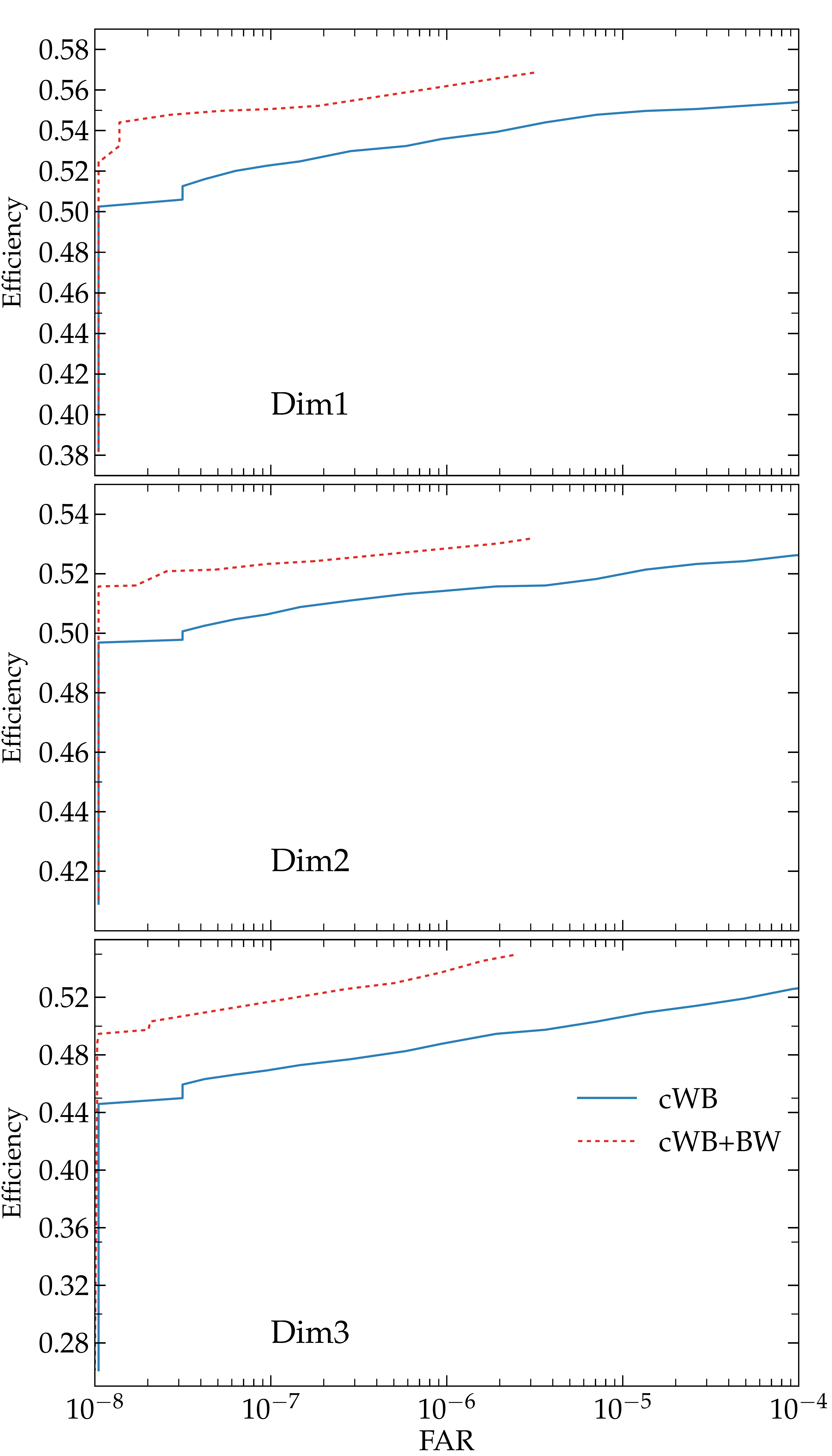}
   \parbox{3.4in}{\caption 	{ROC Improvement using a combination of cWB+BW for \textbf{dim 1/2/3} injected at a nominal distance of 10 kpc with a $50\%$ cWB detection efficiency distance of 0.36 kpc (\textbf{dim1}), 0.42 kpc (\textbf{dim2}), and 1.10 kpc (\textbf{dim3}).\color{black}}}
   \label{Figure 1}
\end{figure}

\begin{figure}[h] 
   \centering
   \includegraphics[width=3.4in]{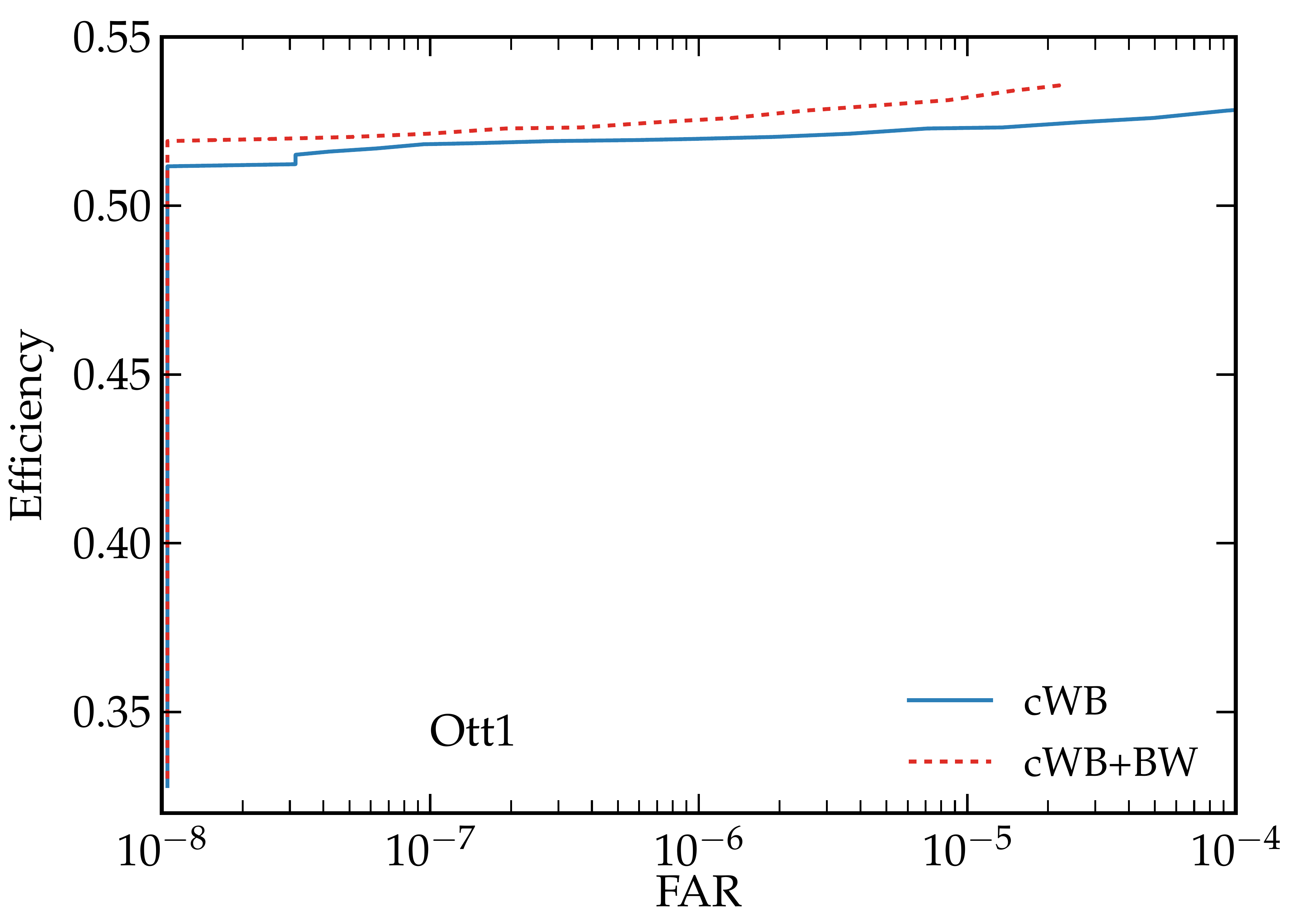}
   \parbox{3.4in}{\caption 	{ROC Improvement using a combination of cWB+BW for \textbf{ott} injected at a nominal distance of 10 kpc with a $50\%$ cWB detection efficiency distance of 0.12 kpc (\textbf{ott}).\color{black}}}
   \label{Figure 1}
\end{figure}

\begin{table*}[t]
\caption{\label{tab:table2}$50\%$ detection efficiency distance increase (using S5 data) shown with cWB+BW ROCs for waveforms possessing both [+, $\times$] polarization components with either the arbitrary or the linearly polarized BW code.}
\begin{ruledtabular}
\begin{tabular}{c|c|c|c|c|c}
Waveform &[$10^{-5}$ FAR]&[$10^{-7}$ FAR]& Avg Arbitrary Polarization & Avg Linear Polarization & $50\%$ Distance (kpc) \\ 
&&&(Improvement) [Volume Increase]&(Improvement)[Volume Increase]&(cWB+BW)\\ [0.5ex] 
\hline 
dim1 & +4.51\%   & +5.69\% & +4.91\% [+15.47\%]& +1.72\% [+5.25\%]& 0.43\\
dim2 & +3.04\%  & +3.31\% & +3.14\% [+ 9.72\%]& +1.18\% [+3.58\%]& 0.48\\
dim3 & +9.17\%  & +9.17\% & +9.59\% [+31.62\%]& +3.19\% [+9.88\%]& 1.40\\
ott &  +1.92\%  & +0.77\% & +1.30\% [+3.95\%]& -4.25\% [-13.30\%]& 2.81\\ 
sch1 & +22.26\%  & +30.72\% & +22.05\% [+81.81\%]& +2.46\% [+7.56\%]& 3.91\\
sch2 & +11.64\%  &  +6.84\% & +9.58\% [+31.58\%]& -19.80\% [-71.94\%]& 2.36\\
sch3 & +1.28\%  & +1.12\% & +1.19\% [+3.61\%]& +0.45\% [+1.36\%]& 1.47\\
\end{tabular}
\end{ruledtabular}
\end{table*}

\begin{table*}[t]
\caption{\label{tab:table2}cWB+BW ROC improvements for waveforms possessing both polarizations from the usage of the arbitrary polarization code.}
\begin{ruledtabular}
\begin{tabular}{c|c|c|c|c|c|c|c}
Waveform & [$10^{-5}$ FAR] & [$10^{-6}$ FAR]  & [$10^{-7}$ FAR] & [$10^{-8}$ FAR] & [$10^{-9}$ FAR] & Avg Arbitrary Improvement & $50\%$ Distance (kpc) \\ 
&&&&&&(Improvement) [Volume Increase]&(cWB+BW)\\[0.5ex] 
\hline 
yak1 & +17.02\% & +19.35\% &  +22.73\% & +23.08\% & +15.38\% & +15.38\% [+53.60\%] & 3.50 \\
yak2 & +15.21\% & +15.53\% &  +15.18\% & +14.29\% & +18.60\% & +15.76\% [+55.12\%]& 4.82\\
yak3 & +8.28\% & +8.54\% &  +7.69\% & +7.85\% & +7.94\% & +8.06\% [+26.18\%]& 3.86\\
yak4 & +3.21\% & +3.66\% &  +2.56\% & N/A & N/A & +3.14\% [+9.72\%]& 5.76 \\[1ex] 
\end{tabular}
\end{ruledtabular}
\end{table*}

\subsubsection{Relevant Polarization States for SN Waveforms}
The linearized Einstein equation in the TT gauge presents wave solutions with two degrees of freedom ($\mathrm{h_{+}}$ and $\mathrm{h_{\times}}$). These degrees of freedom are uncoupled during propagation but coupled at the source through the stress energy tensor and constrained by the fact that the metric is a tensor. It is at the coupling stage that the classifications of different polarizations and the relative amplitude dependence on the source orientation is fixed. Axisymmetric dynamics (realistic for the first few tens of milliseconds of a CCSN from a rapidly rotating progenitor) produce signals that have only one of the two polarizations to be non zero (in an appropriate reference frame). It is important to notice that such source produces no GW at all for an observer along the rotational axis of the  source. Another well studied polarization state is the elliptical one where the two polarizations have a fixed $\pi/2$ phase difference at every frequency,

\begin{equation}
h_\times = \epsilon h_+e^{i\pi /2}
\end{equation}

where $\epsilon$ = [0,1]. 

\indent and that includes the linearly polarized special case when the frequency independent, source orientation dependent, relative factor $\epsilon$ is zero. The case where the $\epsilon=1$ is also named circular polarization. Elliptical polarization (with $\epsilon$ different than zero) is appropriate for some of the extreme phenomenological CCSNe models where the protoneutron star behaves like a rotating ellipsoid\,\cite{2007ApJ...658.1173P}. $\epsilon$ and is given by,

\begin{equation}
\epsilon = \frac{1+cos^2 (\iota)}{(2 cos \iota)}
\end{equation}

and, $\mathrm{\iota}$ is the angle between the line of sight and the rotational axis of the source. Such emission pattern is also automatically enforced from 2D simulations, such as Dimmelmeier\,\cite{2002A&A...393..523D} (of any kind of progenitor). In the case of turbulent, 3D dynamics, such as Ott et al. \,\cite{2013ApJ...768..115O}, none of the approximations above are guaranteed to work and in general it should not be expected, although the difference in the arrival of the two polarizations should be somehow bounded by the fact that are produced at the same time.

\begin{table*}[t]
\caption{Initial set of BW Priors modified.}
\begin{ruledtabular}
\begin{tabular}{ccc}
Priors & Prior Application to CBC Signals & Slowly/Rapidly Rotating CCSNe  \\ [0.5ex] 
\hline 
Sky Location $(\theta, \phi)$ & Uniformly Distributed (All-Sky) & Specific to direction of CCSN \\ [1ex] 
Ellipticity & [0,1] & Set to 0 for linearly polarized CCSNe waveforms \\[1ex]
Polarization of Waveform & Linearly/Circularly/Elliptically polarized & Linear/Arbitrarily polarized\\ [1ex]  
\end{tabular}
\end{ruledtabular}
\end{table*}

\indent At the detector level, a single interferometer is only sensitive to a specific linear combination of the two polarizations and two interferometers are necessary to resolve both polarizations with a quality that depends of the relative misalignment and direction of arrival through the detector responses reviewed in the appendix. For arbitrarily polarized signals, aligned detectors allow to impose stronger consistency constraints and misaligned instruments allow to resolve better the two polarizations. The reconstruction of the direction also improves with the number of interferometers\,\cite{2016PhRvD..93d2004K} but also depends on sky location and relative orientation of the interferometers in the network. For example, cWB can reconstruct the direction of the source in the sky based on coherent network analysis methods\,\cite{2011PhRvD..83j2001K}, which includes both the effects of the delay between different interferometers and the difference in the response. For the two detector network, the time delay of a passing GW between detectors specifies a ring of possible sky locations. cWB calculates likelihood across the ring and the reconstructed sky location is chosen based on a patch in the sky with largest likelihood value. The quantitative assessment of the reconstruction potential (including the direction reconstruction) depends on aspects that can be understood analytically but ultimately requires the numerical studies performed here. 

\indent In order for BW to handle arbitrarily-polarized signals with multi-detector networks, we introduced a separate wavelet basis for the + and $\times$ polarizations. 

\indent Since we do not know the orientation of future CCSNe, we consider the strain for different possible source orientations with respect to the line of sight were projected onto the -2 spin-weighted spherical harmonic basis, $~^{-2}\!Y_{lm}$($\iota, \phi$)\,\cite{2007arXiv0709.0093A}, and average over the results. The effects on the GW polarizations due to the randomization of the source orientation are computed for axisymmetric sources by using polarization factors\,\cite{gossan:15} in order to describe $h_{+,\times}$($\iota, \phi$) as a function of $h_{+,\times;0} = h_{+, \times}$($\iota = 0, \phi = 0$). Defining polarization factors as $n_{+, \times}$($\iota, \phi$), which is dependent on the symmetries of the system in consideration, we write the strain as a function of an arbitrary internal orientation
\begin{equation}
h_{+}(\iota, \phi) = n_+(\iota, \phi)h_{+;0}
\end{equation}
\begin{equation}
h_{\times}(\iota, \phi) = n_\times(\iota, \phi)h_{\times;0}
\end{equation}

In the current analysis, we assume that no knowledge is available on the polarization state of the GW arriving at the detector. Therefore, we test how different tunings perform across the different representative emission models considered. For 2D CCSNe emission models, the axisymmetric system results in a linearly polarized GW signal and we apply 
\begin{equation}
n_+^{lin} = 1
\end{equation}
\begin{equation}
n_{\times}^{lin} = 0
\end{equation}

Hence, we define 
\begin{equation}
h_+(\iota) = h_+^{eq} sin^2 \iota
\end{equation} 
where $h_+^{eq}$ is the strain seen by an equatorial observer. The appropriate SN polarization factors applied would be
\begin{equation}
n_+^{SN} = sin\iota^2
\end{equation}
\begin{equation}
n_{\times}^{SN} = 0
\end{equation}
For 3D CCSNe emission models, no additional polarization factors are applied but the $\mathrm{h_+}$ and $\mathrm{h_\times}$ strains must be computed for specific internal configurations using 
 
\begin{equation}
\mathrm{h_+} - i\mathrm{h_\times} = \frac{1}{D} \sum^{\infty}_{l=2} \sum^{l}_{m=-l} H_{lm}(t)^{-2} \sylm
\end{equation}

where D is the distance and $H_{lm}(t)^{-2}$\,\cite{gossan:15}. Since the quadrupole method of extracting GWs is sufficiently accurate\,\cite{2011PhRvD..83f4008R}, only the $l=2$ mode is considered. 

\begin{figure}[b] 
   \centering
   \includegraphics[width=3.4in]{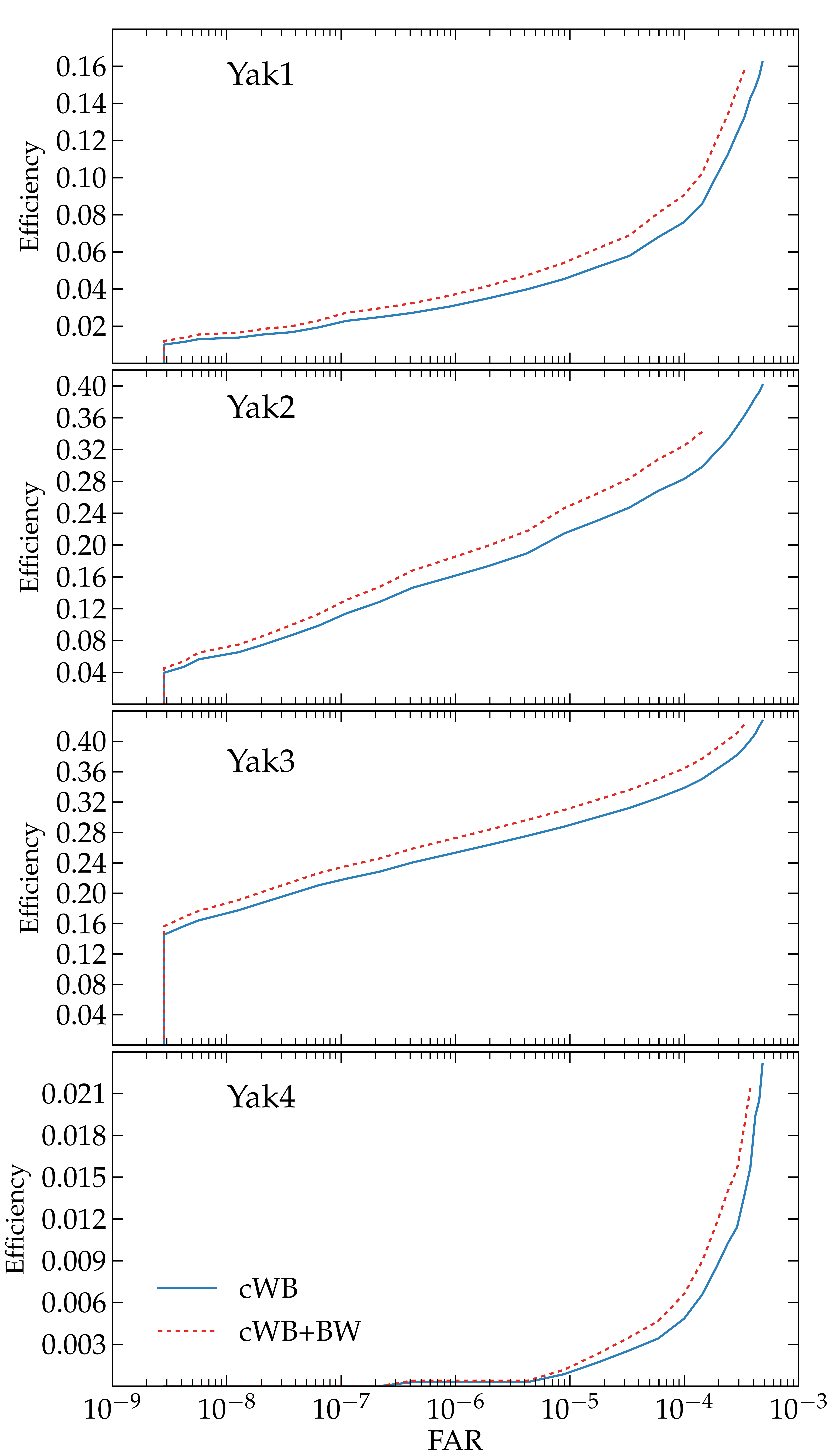}
   \parbox{3.4in}{\caption 	{ROC improvement in the low-FAR parameter space using a combination of cWB+BW for \textbf{yak} injected at a nominal distance of 10 kpc.}}
   \label{Figure 1}
\end{figure}
\subsection{Test of Sky Mask}
\indent In the initial search for CCSNe in the presence of optical triggers with cWB alone\,\cite{2016arXiv160501785A}, it was observed that the search performance was better if we accepted events not only in the correct sky location but also in sky locations within 5$^{\circ}$ degrees of the correct location. This approach to test the reconstruction in unphysical sky locations might seem unintuitive, but it proved beneficial because occasionally the noise was allowing the best reconstruction of the SNR when using a slightly off sky location (equivalent of a phase calibration error in one of the interferometers for a matched filter CBC search). 

\indent We test the same strategy here with the linearly polarized code and analyzed if it allowed a better separation of noise induced events and SN injections. The procedure involved in computing the Bayes factor in the exact location and comparing  them with the calculation for the loudest event in the sky mask in the simulation stage of the analysis. 

\indent The SN waveforms are injected into the detector noise with fixed sky location. The analysis specific to this skymask involves accepting triggers within a circular radial area of 5$^{\circ}$ degrees \,\cite{2016arXiv160501785A} of the SN trigger of interest. Any possible triggers falling outside of this skymask are discarded. We quantify the benefits from the analysis of using a sky mask by observing the shift in mean of the signal-to-glitch ($\mathrm{B_{LN(S/G)}}$) Bayes factor (table V). 

\begin{table*}[h!]
  \centering
  \begin{tabular}{|p{2cm}|c|c|c|c|c|c|c}
    \hline
       {\textbf{Waveform}} & \multicolumn{3}{c|}{cWB + BW} & \multicolumn{3}{c|}{cWB + $\mathrm{BW_{fixed sky}}$} \\
    \cline{2-7}
    & Mean of $\mathrm{B_{LN(S/G)}}$ & STD of $\mathrm{B_{LN(S/G)}}$ & Mean/STD & Mean of $\mathrm{B_{LN(S/G)}}$ & STD of $\mathrm{B_{LN(S/G)}}$ & Mean/STD\\ 
    \hline
    sch1 & 32.25 & 15.07 & 2.14 & 35.87 & 15.58 & 2.30 \\ \hline
    sch2 & -94.59 & 55.70 & -1.70 & -42.62 & 48.13 & -0.89 \\ \hline \color{black}
    sch3 & 19.81 & 26.24 & 0.75 & 34.97 & 23.70 & 1.48 \\ \hline \color{black}
    dim1 & 55.44 & 23.22 & 2.39 & 60.86 & 23.59 & 2.58 \\ \hline \color{black}
    dim2 & 43.49 & 15.53 & 2.80 & 47.58 & 16.32 & 2.92 \\ \hline \color{black}
    dim3 & 17.4 & 12.69 & 1.37 & 19.96 & 13.37 & 1.49 \\ \hline \color{black}
    ott & 15.82 & 38.50 & 0.41 & 26.22 & 34.52 & 0.76 \\ [1ex] \hline
  \end{tabular}
  \caption{Quantified improvement (through the analysis of the shift in mean) of the signal-to-glitch ($ln_{SG}$) Bayes factor classification with the implementation of a sky-location prior. The mean shift, with the implementation of the sky mask, seems to consistently indicate that the injections are more likely to be confused with noise and therefore making it harder to distinguish signals from glitches.}
\end{table*}

\indent Similarly to the previous section, since all possible source inclination angles are equally likely, we inject CCSNe signals with many variations of ($\iota, \phi$)\,\cite{2016PhRvD..93d2002G}. cWB reconstructs the direction of the source in the sky based on coherent network analysis methods\,\cite{2011PhRvD..83j2001K}, which includes both the effects of the delay between different interferometers and the difference in the response. The performance metric was evaluated after we compensate by injecting CCSNe signals with many different angle variations of ($\iota, \phi$) in order to average over all internal source orientations\,\cite{2016PhRvD..93d2002G}. 

\begin{figure}[t] 
   \centering
   \includegraphics[width=3.4in]{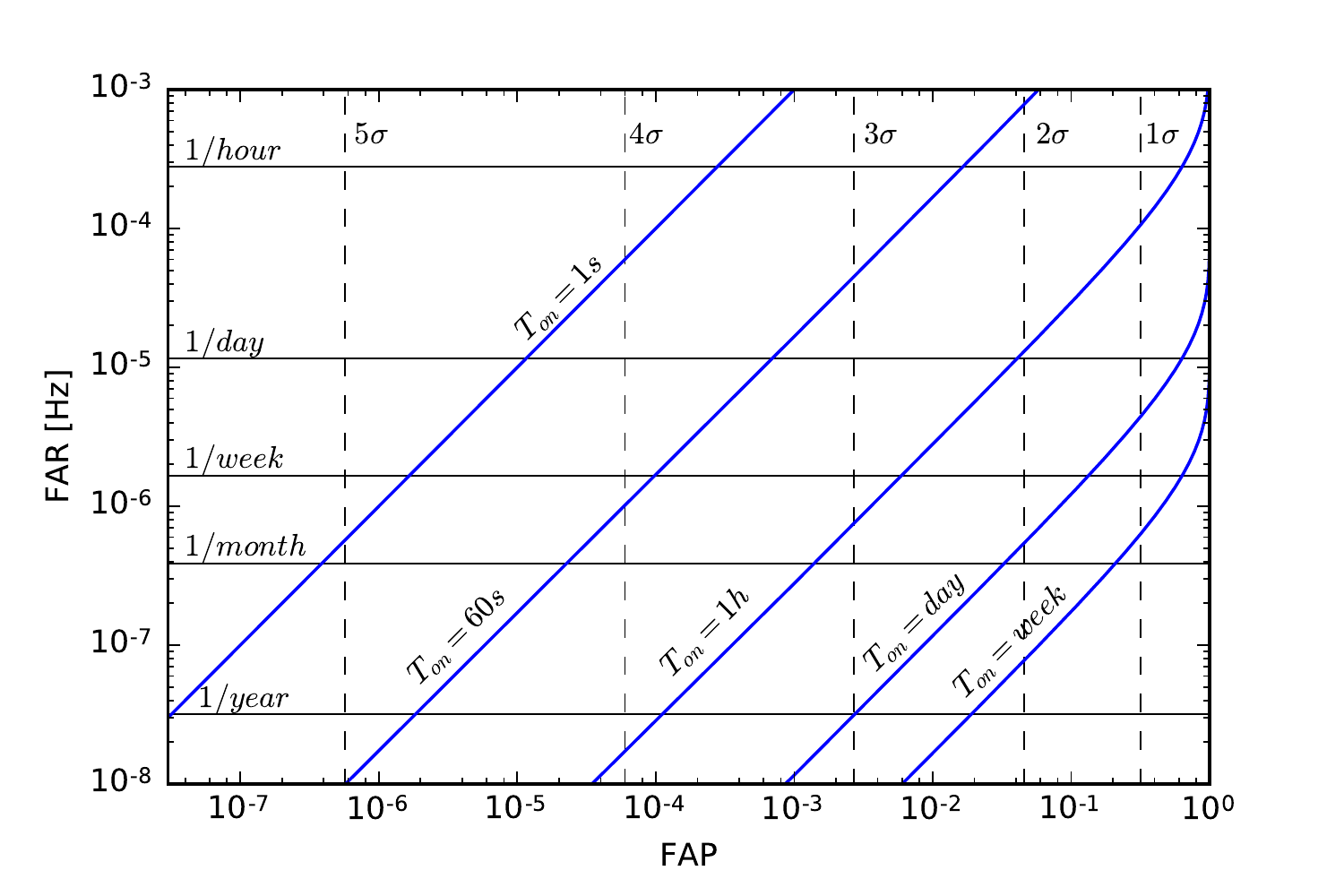}
   \parbox{3.4in}{\caption 	{For applying the results of this paper to different on-source windows, this plot illustrates that for any desired number of signals or a FAP with a specific on-source window duration the resultant FAR values expected for the accompanying ROC curve. The detection scenario with a $3\sigma$ confidence is interesting, particularly for on-source windows of: 60s, 1hr, 1 day, and 1 week with corresponding FAR ranges of: $4.803\times10^{-5}$, $7.630\times10^{-7}$, $3.363\times10^{-8}$, and $3.299\times10^{-8}$.}}
   \label{Figure 1}
\end{figure}

\section{Discussion}

\indent We study the performance of BW, a Bayesian GW burst parameter estimation follow up module, with the purpose of discriminating CCSNe related events from noise induced events. Past applications of BW\,\cite{2015CQGra..32m5012C, 2016PhRvD..94d4050L, 2016PhRvD..93b2002K, 2016arXiv161202003B} have demonstrated promise across CBC signals where a large fraction of the GW energy is in the merger phase.

\indent Two versions of the BW code were tested as follow up to cWB: (a) a pre-existing one that assumed elliptical polarization of the signals (with a tunable ellipticity parameter), and (b) a new one that does not make any assumptions on the polarization state of the signal.

\indent The performance of the preexisting code was not consistent among GW morphologies, meaning that for linearly polarized signal it gave comparable results to the new code, but for GW SN signals arbitrarily polarized the performance was worse than cWB alone.

\indent The summary of this work is that the code developed for arbitrary polarizations performs more robustly than the code developed for generic burst transients. We therefore propose to use the arbitrary polarization code in future CCSN searches.

\indent The best way to separate GW events from noise induced events appeared to be

\begin{equation}
\mathrm{ln(signal/noise)} = 2.04323 - 67.4947\mathrm{[ln(signal/glitch)]}
\end{equation}

instead of the previously used ln(signal/noise)>0 and ln(signal/glitch)>0 (see Fig 9 to 12).


\indent The cWB+BW ROC range improvements with respect to a fixed FAR for slowly-rotating waveforms were: $+1.30\%$ for \textbf{ott}, $+15.38\%$ for \textbf{yak1}, $+15.76\%$ for \textbf{yak2}, $+8.06\%$ for \textbf{yak3}, $+3.14\%$ for \textbf{yak4}, while for rapidly-rotating waveforms were: $+4.91\%$ for \textbf{dim1}, $+3.14\%$ for \textbf{dim2}, $+9.59\%$ for \textbf{dim3}, $+22.05\%$ for \textbf{sch1}, $+9.58\%$ for \textbf{sch2}, $+1.19\%$ for \textbf{sch3} (tables II-III and fig 3-7). 

\indent In general, for sources with SNR in the so called "Gaussian wall region" (where the rate decreases quickly with $\rho$), BW is expected to help with the improvement of the statistical confidence and the ROCs are expected to be fairly flat. For example, as shown in figure 6, the FAR is reduced by more than one order of magnitude without changing significantly the efficiency (it can be observed because in this case the operating points cWB moved roughly horizontal to the left to the cWB+BW ROC curve). If there was the presence of a strong glitch background, the ROC curves would have been more sloped (because the FAR tends to change more slowly with the SNR) and there would have been a greater improvement in the range.

\indent The general way to interpret the improvement in the cWB+BW ROC vs the cWB ROCs displayed in figures 3-7 is twofold: (a) the vertical displacement of an operating point describes the improvement in the efficiency or equivalently range and visible volume at a specific FAP. The second part can be derived from past cWB efficiency curves and see what shift in the values of the detection ranges produces the improvement in the detection efficiency observed in the plots produced here. (b) The location of the derived operating for cWB+BW horizontally to the left of the cWB operating point quantify the benefit in detection confidence. A summary of the benefits in terms of FAR reduction and range increase for the scenarios explored in this paper is displayed in tables II and III.

\section{Appendix}
\subsection{Detector Responses}
For the two detector network, the time delay of a passing GW between detectors specifies a ring of possible sky locations. cWB calculates likelihood across the ring and the reconstructed sky location is chosen based on a patch in the sky with largest likelihood value.

The measurements, from a network of $N$ detectors, are defined by

\begin{equation}
x = \mathrm{Fh + e} 
\end{equation}

where x is the vector of measurements $[x_1,..., x_N]^T$, the matrix F = $[[F_1^+, F_1^\times],...,[F_N^+, F_N^\times]]^T$ contains the antenna responses of the observatories to the GW strain vector, h = $[h_+, h_\times]^T$, and $e$ is the noise in each sample. The antenna patterns matrix, F, is a known function of the source sky direction, $(\theta, \phi)$, and the decomposition into + and $\times$ polarizations depends on the polarization basis angle $\psi$.

\indent A GW signal is characterized by a set of five angles: ($\theta, \Phi, \psi$), that describe the sky location and polarization basis, while ($\iota, \phi$) describe the internal orientation of the source relative to the observer's line of sight. The strain detected by a GW interferometer, h(t), is given by,

\begin{equation}
h(t) = F_+(\theta, \Phi, \psi) h_+(t) + F_{\times} (\theta, \Phi, \psi) h_{\times}(t)
\end{equation}

where $F_{+, \times}$ ($\theta, \Phi, \psi$) are the antenna response functions of the detector to the two GW polarizations, $h_{+, \times}(t)$. $F_{+, \times}$ are given by,

\begin{equation}
F_+ = \frac{1}{2} (1+\mathrm{cos}^2\theta)\mathrm{cos}2\phi \mathrm{cos}2\psi - \mathrm{cos}\theta \mathrm{sin}2\phi \mathrm{sin}2\psi
\end{equation}

\begin{equation}
F_{\times} = \frac{1}{2} (1+\mathrm{cos}^2\theta)\mathrm{cos}2\phi \mathrm{sin}2\psi - \mathrm{cos}\theta \mathrm{sin}2\phi \mathrm{cos}2\psi
\end{equation}

We generalize the analysis to the case of CCSNe signals of unknown source sky direction, $(\theta, \phi)$, and arrival time, $\tau$, with respect to the center of the Earth in the observation window of $\mathrm{f_s}^{-1}$M seconds with priors being defined as 

\begin{equation}
p(\theta|H_1) = \frac{1}{2}\mathrm{sin}(\theta)
\end{equation}

\begin{equation}
p(\phi|H_1) = (2\pi)^{-1}
\end{equation}

\begin{equation}
p(\tau|H_1) = \mathrm{f_s}\mathrm{M}^{-1}
\end{equation}

A global network of $N$ GW detectors each produce a time-series of $M$ observations with sample frequency, $\mathrm{f_s}$, with each of the responses being instead defined as\,\cite{2009CQGra..26o5017S}

\begin{equation}
h(t - \Delta(t)) = F_+(\theta, \Phi, \psi) h_+(t - \Delta(t)) + F_{\times} (\theta, \Phi, \psi) h_{\times}(t - \Delta(t))
\end{equation}

\section{Acknowledgements}
This paper was reviewed by the LIGO Scientific Collaboration under LIGO Document P1600353. The authors thank Neil Cornish, Jonah Kanner, Sergey Klimenko, Tyson Littenberg, Meg Millhouse, and Salvatore Vitale for the insightful discussions and their valuable comments on the manuscript. The authors acknowledge the support of the National Science Foundation, the LIGO Laboratory, and NSF grant "NSF PHY-1505861" (PI's: Romano, Mukherjee and Mohanty). LIGO was constructed by the California Institute of Technology and Massachusetts Institute of Technology with funding from the National Science Foundation and operates under cooperative agreement PHY-0757058. The authors would like to acknowledge the use of the LIGO Data Grid computer clusters for performing all the computations reported in the paper.

\bibliographystyle{apsrev}
\bibliography{BW_arxiv}
\clearpage
\begin{figure}[h] 
   \centering
   \includegraphics[width=7in]{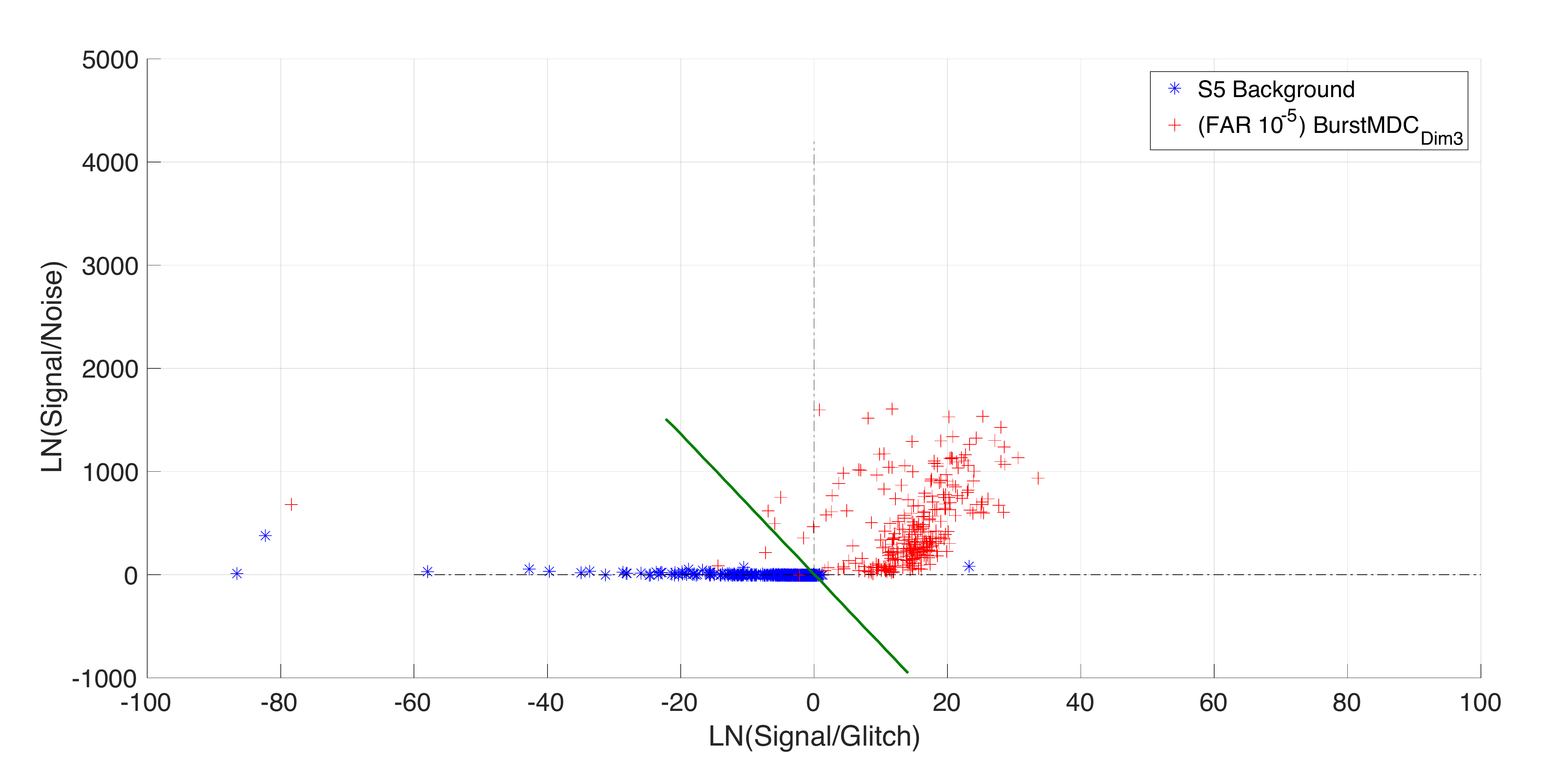}
   \parbox{7in}{\caption 	{The Bayes factor scatterplot of \textit{ln(Signal/Glitch)} vs \textit{ln(Signal/Noise)} for noise events produced with 3.71 days of H1/L1 data with a FAR of $10^{-5}$ and SN induced events with the rapidly rotating \textbf{dim3} model injected at a nominal distance of 10 kpc.}}
   \label{Figure 1}
\end{figure}
\clearpage
\begin{figure}[h] 
   \centering
   \includegraphics[width=7in]{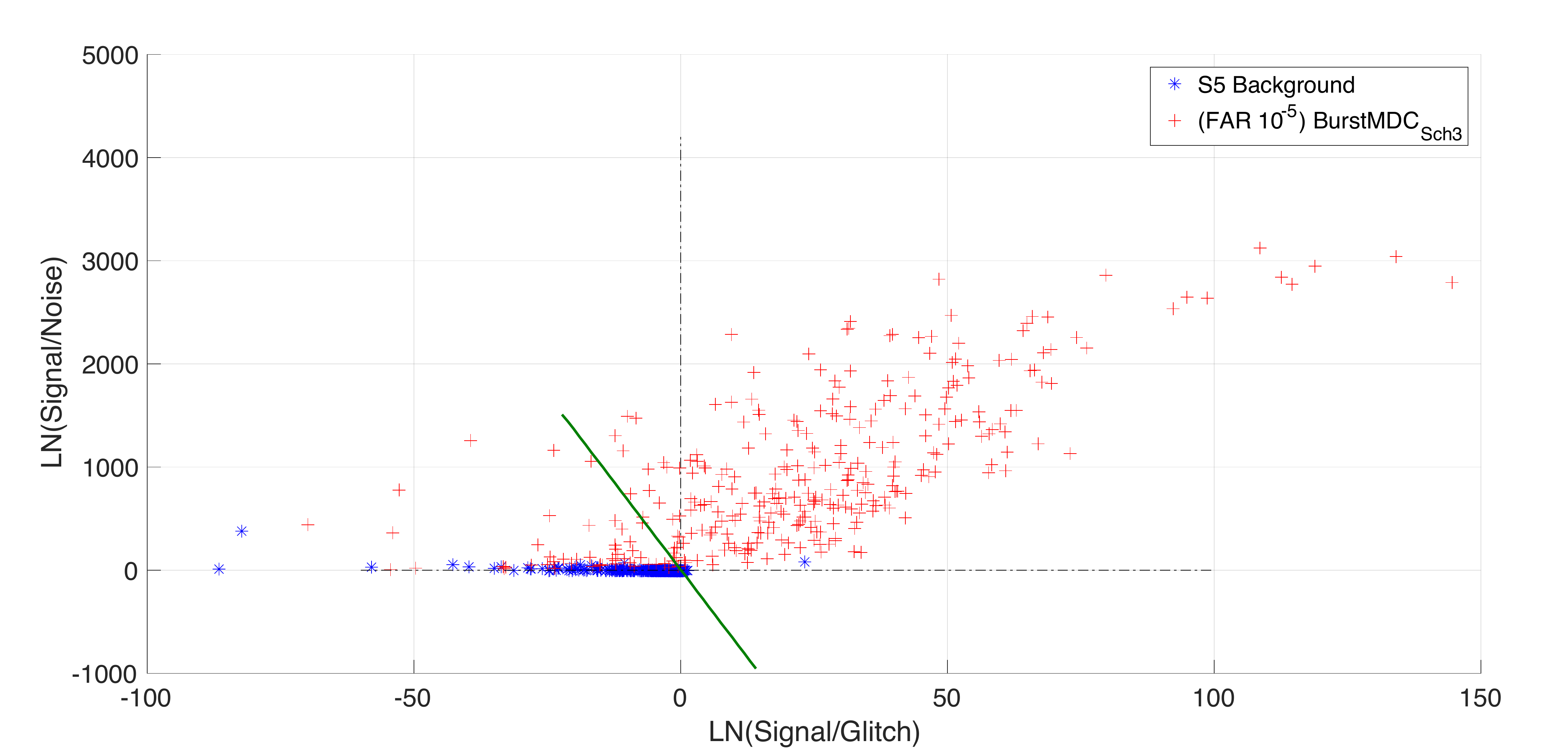}
   \parbox{7in}{\caption 	{The Bayes factor scatterplot of \textit{ln(Signal/Glitch)} vs \textit{ln(Signal/Noise)} for noise events produced with 3.71 days of H1/L1 data with a FAR of $10^{-5}$ and SN induced events with the rapidly rotating \textbf{sch3} model injected at a nominal distance of 10 kpc.}}
   \label{Figure 2}
\end{figure}
\clearpage
\begin{figure}[t] 
   \centering
   \includegraphics[width=7in]{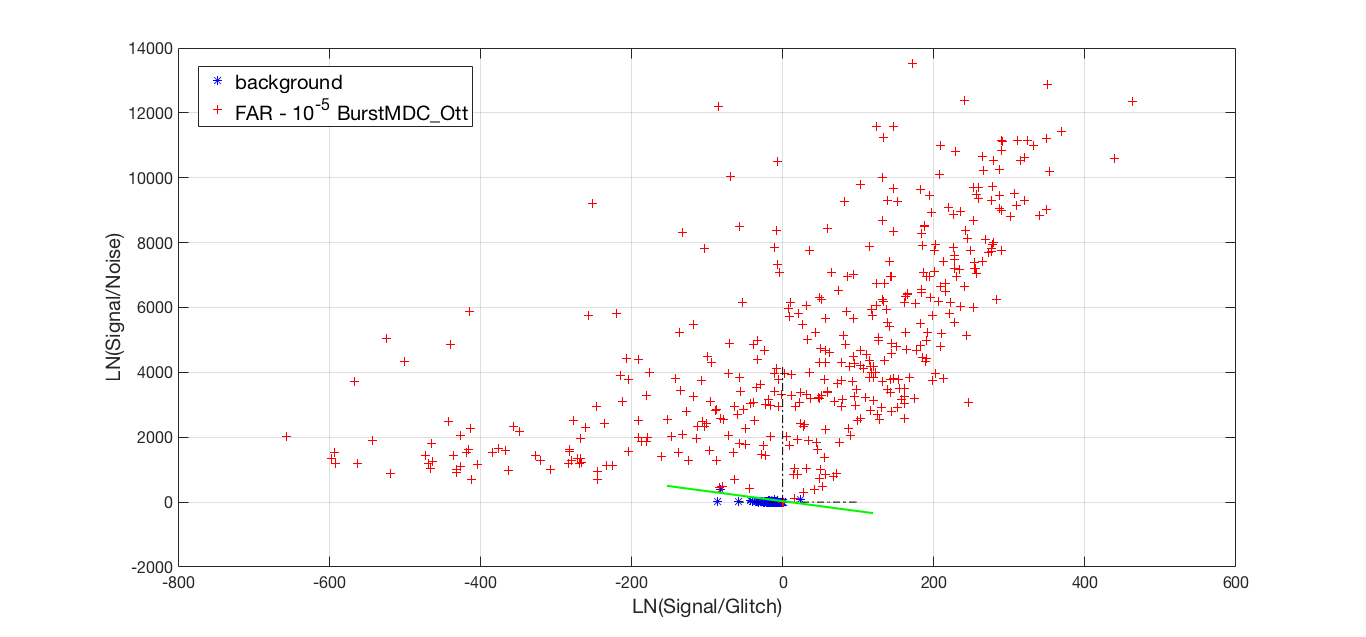}
   \parbox{7in}{\caption 	{The Bayes factor scatterplot of \textit{ln(Signal/Glitch)} vs \textit{ln(Signal/Noise)} for noise events produced with 3.71 days of H1/L1 data with a FAR of $10^{-5}$ and SN induced events with the non-rotating \textbf{ott} model. The poor separation between the populations of noise and GW events populations resulted in a relaxed slope (green line) for the separation line used in the tuning for \textbf{ott}.}}
   \label{Figure 1}
\end{figure}
\clearpage
\begin{figure}[t] 
   \centering
   \includegraphics[width=7in]{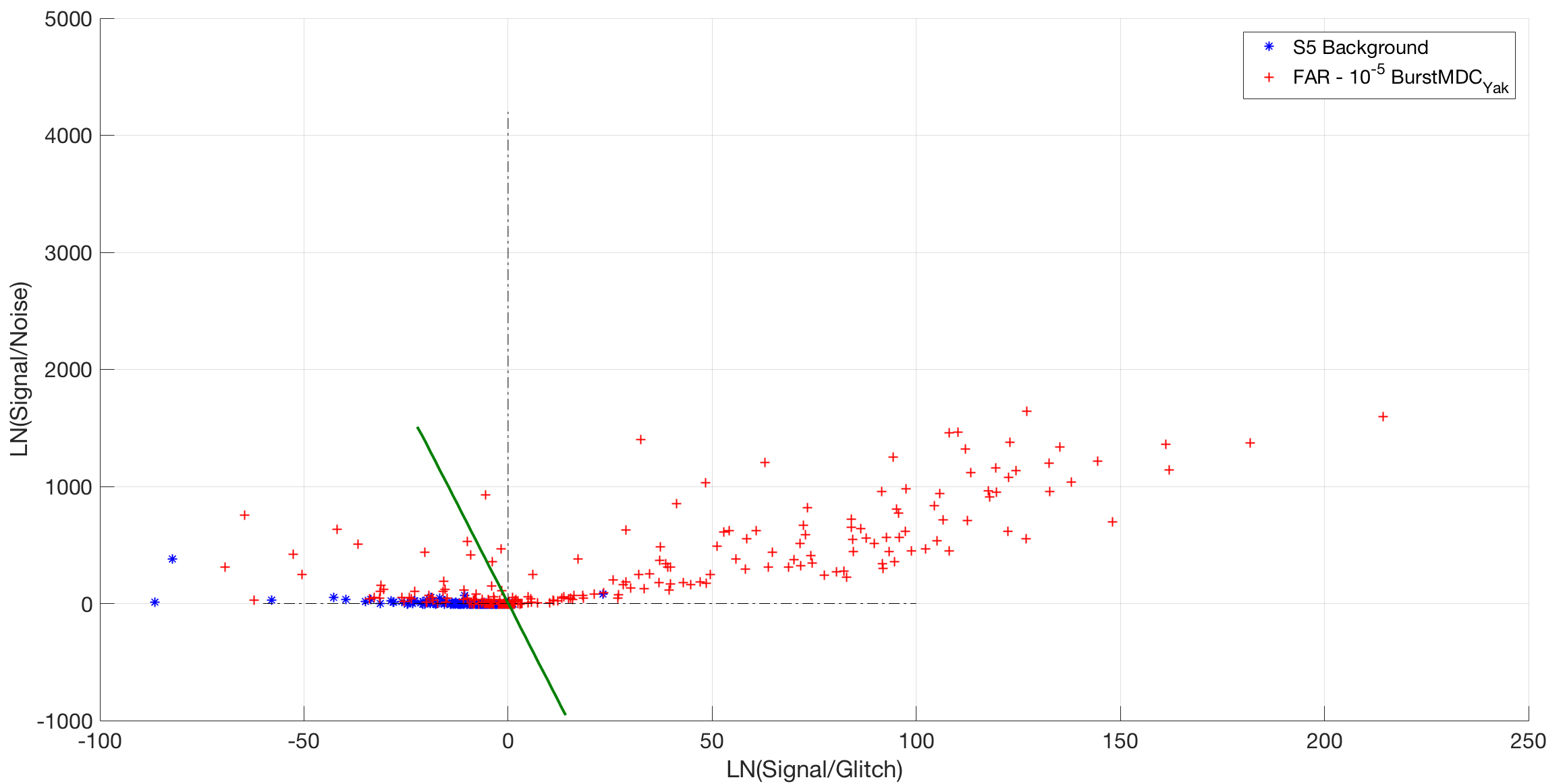}
   \parbox{7in}{\caption 	{The Bayes factor scatterplot of \textit{ln(Signal/Glitch)} vs \textit{ln(Signal/Noise)} for noise events produced with 3.71 days of H1/L1 data with a FAR of $10^{-5}$ and SN induced events with the non-rotating \textbf{yak} model injected at a nominal distance of 10 kpc.}}
   \label{Figure 1}
\end{figure}

\end{document}